\RequirePackage{fix-cm}
\documentclass[amsthm,smallextended]{svjour3}       
\smartqed  
\usepackage{graphicx}
\usepackage[round]{natbib}
\usepackage{amsmath}
\usepackage{amssymb}
\usepackage{amsfonts}
\usepackage[ruled]{algorithm2e}
\usepackage{dcolumn}
\usepackage{tikz}
\usepackage{booktabs}
\usepackage{hyperref}
\usepackage[detect-all]{siunitx}
\usepackage{array,multirow}
\usepackage{bm}

\makeatletter
\newcolumntype{B}[3]{>{\boldmath\DC@{#1}{#2}{#3}}c<{\DC@end}}
\makeatother

\newcommand{\rebut}[1]{#1}
\newcommand{\rebutt}[1]{#1}

\begin{document}

\title{On the Parenthesisations of Matrix Chains
}
\subtitle{All Are Useful, Few Are Essential}


\author{Francisco López         \and
        Lars Karlsson           \and
        Paolo Bientinesi
}

\authorrunning{F. López, L. Karlsson, P. Bientinesi} 

\institute{Francisco López, Lars Karlsson, Paolo Bientinesi \at
              Department of Computing Science, Umeå Universitet, Sweden \\
              \email{\{flopz, larsk, pauldj\}@cs.umu.se}           
}

\date{Received: date / Accepted: date}

\maketitle

\begin{abstract}
The product of a matrix chain consisting of $n$ matrices can be computed in $C_{n-1}$ (Catalan's number) different ways, each identified by a distinct parenthesisation of the chain. 
The best algorithm to select a parenthesisation that minimises the cost runs in $O(n \log n)$ time. 
Approximate algorithms run in $O(n)$ time and find solutions that are guaranteed to be within a certain factor from optimal; the best factor is currently $1.155$. 
In this article, we first prove two results that characterise different parenthesisations, and then use those results to improve on the best known approximation algorithms. 
Specifically, we show that (a) each parenthesisation is optimal somewhere in the problem domain, and (b) exactly $n + 1$ parenthesisations are essential in the sense that the removal of any one of them causes an unbounded penalty for an infinite number of problem instances. 
By focusing on essential parenthesisations, we improve on the best known approximation algorithm and show that the approximation factor is at most $1.143$. 


\end{abstract}

\section{Introduction}\label{sec::intro}

We consider the classical matrix chain multiplication problem (MCP), where the cost of multiplying a $p \times q$ matrix by a $q \times r$ matrix is taken to be $p q r$.
\begin{problem}[MCP]
    \label{problem:matrix-chain}
    Given $n \ge 3$, let 
    $\boldsymbol{k} = (k_{0}, k_{1}, \dots, k_{n})$ be a sequence of $n+1$ positive integers, 
    and for each $i \in \{ 1, 2,\ldots,n\}$, let $M_i$ be a matrix of size $k_{i-1}\times k_i$. Given the assignment 
    \begin{equation}
        \label{eq:mc-n}
        M := M_1 M_2 \cdots M_n,
    \end{equation}
    which is normally referred to as a \emph{matrix chain}, find a parenthesisation that minimises the cost of computing $M$.
    \rebut{Oftentimes, this cost captures the number of multiplications (and possibly additions) to be performed.
    The parenthesisation that minimises this cost is said to be \emph{optimal}.}
    If the optimal parenthesisation is unique, it is said to be \emph{uniquely optimal}.
    The integer sequence $\boldsymbol{k}$ specifies an instance of the problem.
    \qed
\end{problem}

Because of the associativity of matrix multiplication, 
a matrix chain of length $n$ can be computed in  
$C_{n-1} = \frac{(2n-2)!}{n!(n-1)!}$ different ways, known as "parenthesisations" ($C_{n}$ is the $n$-th Catalan number); 
different parenthesisations may have different costs. 

The MCP can be solved exactly in $\Theta(n^3)$ time using Godbole's classical dynamic programming algorithm from~\cite{godbole1973efficient}.
The fastest exact algorithms run in $O(n \log n)$ time, given in~\cite{hu1982computation,hu1984computation}, 
and the fastest approximation algorithms run in $O(n)$ time, presented in~\cite{chandra1975computing} and~\cite{chin1978n}, with approximation factors smaller than $2$.

The MCP is a discrete optimisation problem over the set of all parenthesisations. 
One of the aims of this paper is to investigate the nature of these parenthesisations. 
In particular, we show that every parenthesisation is \emph{useful} in the sense that each is uniquely optimal in an infinite subset of the problem domain. 
Therefore, any algorithm that solves the MCP exactly must consider every parenthesisation; none of them are superfluous. 
\rebut{We also show that, when the problem is solved approximately, some parenthesisations are more important than others. In fact, some of the  parenthesisations can be ignored without causing the approximation error to grow above a factor of two. By contrast, other parenthesisations, which we refer to as \emph{essential}, when removed make the approximation error unbounded. These concepts are properly defined in the following sections. }

Armed with these new concepts, we re-interpret Chandra's and Chin's approximation algorithms and propose modifications to both that slightly improve the approximation without negatively impacting the time complexity.

\section{Related Work}\label{sec::related_work}
The MCP has been studied extensively; many exact and approximate algorithms exist.  
We begin by outlining the history of the exact algorithms.
In 1973,~\cite{godbole1973efficient} presented a dynamic programming algorithm with $\Theta(n^3)$ time complexity. 
Godbole's algorithm has since been used as a textbook example of dynamic programming.
In 1982,~\cite{yao1982speed} presented an improved dynamic programming algorithm with $O(n^2)$ time complexity. 
In a two-part paper published in 1982 and 1984,~\cite{hu1982computation,hu1984computation} presented an exact algorithm with $O(n \log n)$ time complexity, which remains the fastest known exact algorithm. 
The problem complexity is conjectured to be $\Omega(n \log n)$, but this has not yet been proven~\cite{Ramanan1994}. 
In 2019, a flaw in the proof of Lemma~1 in~\cite{hu1984computation} was found and corrected by~\cite{schwartz2019revisiting}.
The work by Hu and Shing is significant for other reasons as well. 
They showed that there is a one-to-one correspondence between the MCP with $n$ matrices and the problem of triangulating a convex polygon with $n + 1$ vertices in~\cite{hu1982computation}. 
The matrices translate into edges, the matrix dimensions into vertex weights, and the multiplications into triangles. 
The cost of a triangle is the product of the three vertices' weights, and the cost of a triangulation is the sum of the costs of its triangles. 

Approximation algorithms began to appear soon after~\cite{godbole1973efficient} formulated the MCP.
Two years later, in 1975,~\cite{chandra1975computing} presented a $\Theta(n)$ algorithm and proved its approximation factor to be $2$ and that it cannot be improved.
Sadly, Chandra's work was never published other than as an IBM Research technical report.
Soon thereafter, in 1978,~\cite{chin1978n} presented a more accurate $\Theta(n)$ approximation algorithm and proved that the approximation factor is at most $1.25$.
Experiments led Chin to conjecture that the bound could be tightened to $1.155$.
In 1981,~\cite{hu1981n} used Chin's ideas to formulate an approximation algorithm to the equivalent polygon triangulation problem.
They simultaneously proved that their algorithm (and hence Chin's) has an approximation factor of $1.155$ and that this bound is tight; thus validating Chin's conjecture.

\section{Preliminaries}\label{sec::preliminaries}

In this section, we enumerate some definitions and basic results that are used throughout the text. 
Let $\mathcal S_n$ denote the set of parenthesisations for a matrix chain \eqref{eq:mc-n} with $n$ matrices, and let 
$\boldsymbol{k} = (k_0, k_1, \ldots, k_n)$ be an instance of the MCP.
The \emph{cost} of multiplying the partial product $(M_{a+1} \cdots M_b)$ with the partial product $(M_{b+1} \cdots M_c)$ is denoted by
\begin{equation}
    \label{eq:cost-function-phi}
    \phi_{\boldsymbol k}(a, b, c) := k_a k_b k_c,
\end{equation}
where $0 \leq a < b < c \leq n$. 
More general cost functions have been considered in~\cite{chandra1975computing}.

A matrix chain of length~$n$ is evaluated by $n - 1$ multiplications. 
A parenthesisation specifies a partial ordering of these multiplications and each multiplication can be uniquely identified by a triplet $(a, b, c)$ as described above. 
Therefore, a parenthesisation can be uniquely identified by a set $\mathcal A$ of $n - 1$ triplets $(a, b, c)$. 
The cost of a parenthesisation $\mathcal A$ on an instance $\boldsymbol{k}$ can hence be expressed as
\begin{equation*}
    T(\mathcal{A},\boldsymbol{k}) = \sum_{(a, b, c) \in \mathcal{A}} \phi_{\boldsymbol k} (a, b, c) = \sum_{(a,b,c)\in \mathcal A} k_a k_b k_c.
\end{equation*}
For example, consider the parenthesisation $((M_{1} M_{2}) M_{3}) M_{4}$, which is identified by the triplets $\mathcal A = \{(0,1,2), (0,2,3), (0,3,4)\}$.
Thus, the cost is $T(\mathcal A, \boldsymbol k) = k_0 k_1 k_2 + k_0 k_2 k_3 + k_0 k_3 k_4$. 

Recall that the MCP is an optimisation over all parenthesisations $\mathcal S_n$. 
If a subset $\mathcal Q \subset \mathcal S_n$ is excluded from the optimisation, then the optimal cost on some problem instances may increase as a result.
We introduce two functions that quantify this increase on (a) a specified problem instance and (b) the entire problem domain.
The function
\begin{equation*}
    P(\mathcal{Q}, \boldsymbol{k}) :=  
    \frac{
        \min_{
            \mathcal{B} \in \mathcal{S}_n \setminus \mathcal{Q}
        }
        T(\mathcal{B}, \boldsymbol{k})
    }{
        \min_{
            \mathcal{A} \in \mathcal{S}_n
        }
        T(\mathcal{A}, \boldsymbol{k})
    } 
    - 1 
\end{equation*}
measures the relative increase in optimal cost when the parenthesisations in $\mathcal Q$ are removed. 
\rebut{We refer to $P(\mathcal{Q}, \boldsymbol{k})$, which lies in the range $[0, \infty)$, as the \emph{penalty} on instance $\boldsymbol{k}$ caused by removing $\mathcal Q$.}
For example, if the penalty is $0.3$, then the optimal cost increased by $30\%$ on the instance $\boldsymbol{k}$ as a result of removing $\mathcal Q$.
The penalty is zero if and only if $\mathcal S_n \setminus \mathcal Q$ contains an optimal parenthesisation for the given instance.
The following function measures the \emph{maximum penalty} over all possible instances:

\begin{equation}\label{eq:max_penalty_set}
    P(\mathcal{Q}) := \sup_{\boldsymbol{k} \in \mathbb{N}_{+}^{n+1}} P(\mathcal Q, \boldsymbol k).
\end{equation}

The following result from~\cite{chin1978n} is central to several of our proofs.
\rebut{
Informally, if a matrix chain contains a pair of adjacent matrices $M_i \times M_{i+1}$ that look sufficiently like a block inner product ($k_i \gg k_{i-1}, k_{i+1}$), then they must be multiplied together in any and all optimal parenthesisations.  
}%
We restate the result here in a slightly restricted form. 
\begin{lemma}
    \label{thm:chin}
    Let $\boldsymbol k = (k_0, k_1, \ldots, k_n)$ be an instance and let $k_m = \min_i k_i$.
    If for some $i \in \{1, 2, \ldots, n - 1\}$, $k_i > k_{i-1}$ and $k_i > k_{i+1}$ and the inequality 
    \begin{equation*}
        \frac{1}{k_i} < \frac{1}{k_{i-1}} + \frac{1}{k_{i+1}} - \frac{1}{k_m}
    \end{equation*}
    is satisfied, then the multiplication $M_i \times M_{i+1}$ must appear in every optimal parenthesisation.
\end{lemma}
\begin{proof}
    See the proof of~\cite[Theorem~1]{chin1978n}. \qed
\end{proof}

\section{All parenthesisations are useful}\label{sec:algorithms_useful}

A matrix chain with $n$ matrices has $C_{n-1}$ distinct parenthesisations. 
In this section, we answer the question of whether or not all of them are \emph{useful}. 
Equivalently, we determine if some of them can be discarded without increasing the optimal cost anywhere in the problem domain. 

\begin{definition}[Useful parenthesisation]\label{def:useful}
A parenthesisation $\mathcal A$ is \emph{useful} if there exists an instance $\boldsymbol{k}$ on which it is uniquely optimal.
That is, $P(\{ \mathcal A \}, \boldsymbol{k}) > 0$ and, therefore, $P(\{ \mathcal A \}) > 0$.
\end{definition}

The following lemma shows that any useful parenthesisation is uniquely optimal not just on one or a finite number of instances but on an infinitely large subset of the problem domain.
\rebut{
The proof proceeds by constructing a ray from an instance on which a parenthesisation is uniquely optimal. 
That parenthesisation remains uniquely optimal on all the infinitely many instances on the ray. 
}

\begin{lemma}\label{lemma::infinity}
A parenthesisation that is uniquely optimal on one instance is uniquely optimal on infinitely many instances.
\end{lemma}
\begin{proof}
Let $\mathcal A$ be a parenthesisation that is uniquely optimal on instance $\boldsymbol{k} = (k_{0}, k_{1}, \ldots, k_{n})$.
For all other parenthesisations $\mathcal B \neq \mathcal A$, we have by definition
\begin{displaymath}
    \frac{T(\mathcal{B},\boldsymbol{k})}{T(\mathcal{A},\boldsymbol{k})} > 1.
\end{displaymath}
The instance $\boldsymbol{k}$ can be scaled by any integer $\alpha > 1$:
\begin{displaymath}
\boldsymbol{k}' = \alpha \boldsymbol{k} = (\alpha k_0, \alpha k_1, \ldots, \alpha k_n).    
\end{displaymath}
Then $\phi_{\boldsymbol{k}^{'}}(a, b, c) = (\alpha k_a) (\alpha k_b) (\alpha k_c) = \alpha^{3} \phi_{\boldsymbol{k}}(a, b, c)$
from which it follows that
\begin{displaymath}
\frac{T(\mathcal{B},\boldsymbol{k}')}{T(\mathcal{A},\boldsymbol{k}')} = \frac{\alpha^{3}T(\mathcal{B},\boldsymbol{k})}{\alpha^{3}T(\mathcal{A},\boldsymbol{k})} =
\frac{T(\mathcal{B},\boldsymbol{k})}{T(\mathcal{A},\boldsymbol{k})}  > 1.
\end{displaymath}
Since this holds for any parenthesisation $\mathcal B \neq \mathcal A$, we can conclude that $\mathcal{A}$ is uniquely optimal also on instance $\boldsymbol{k}'$ and that the ratio between their costs remains the same.
Since there are infinitely many choices for $\alpha > 1$, each resulting in a distinct instance, there are infinitely many instances on which $\mathcal{A}$ is uniquely optimal. \qed
\end{proof}

\rebut{
The following result states that for every parenthesisation we can find infinitely many instances where that parenthesisation has a strictly lower cost than all others.
\rebutt{
In other words, every parenthesisation is useful; no parenthesisation is redundant.
This implies that no parenthesisation $\mathcal{A}$ can be removed from $\mathcal{S}_n$ without making $P(\{\mathcal{A}\}) > 0$.
}
Therefore, any algorithm that solves the MCP exactly must be able to output every possible parenthesisation.
\rebutt{
The proof of Theorem~\ref{theorem::unique_optimal} is by construction of an instance on which a given parenthesisation is uniquely optimal.
We do so by systematically assigning dimensions such that the repeated application of Lemma~\ref{thm:chin} leads to the desired parenthesisation being optimal.
Finally, we apply Lemma~\ref{lemma::infinity} to show that the given parenthesisation is uniquely optimal on infinitely many instances.
} 
} 

\begin{theorem}\label{theorem::unique_optimal}
Every parenthesisation is uniquely optimal on infinitely many instances. 
\end{theorem}

\begin{proof}
The case $n \leq 2$ is trivial, so assume $n \geq 3$.
Consider an arbitrary parenthesisation $\mathcal A \in \mathcal S_n$.
We will construct an instance $\boldsymbol{k}$ on which $\mathcal{A}$ is uniquely optimal and hence useful per Definition~\ref{def:useful}.
We construct the instance in two steps.
In Step~1, we craft an increasing sequence of positive integers.
From this sequence, in Step~2 we construct an instance $\boldsymbol{k}$ on which  
$\mathcal A$ is uniquely optimal. 
Finally, we apply Lemma~\ref{lemma::infinity} to go from one to infinitely many instances. 
The result then follows.

\emph{Step 1. Craft an increasing sequence of positive integers.}
Let $n$ be the length of the matrix chain.
Let $1 = R_0 < R_1 < \ldots < R_{n-1} = 2$ be the increasing sequence of positive rational numbers defined by
\begin{displaymath}
    R_{\ell} = \frac{ 2^n + \ell - n - 1 } { 2^n + \ell - n - 2^\ell }, \quad \ell \in \{0, 1, \ldots, n-1\}.
\end{displaymath}
If $\ell_1, \ell_2 < \ell$, then
\begin{equation}
    \label{eq:thm:unique_optimal:ineq}
    \frac{1}{R_\ell} < \frac{1}{R_{\ell_1}} + \frac{1}{R_{\ell_2}} - \frac{1}{R_0}.
\end{equation}
To verify this, note that $R_0 = 1$ and $R_{\ell_1}, R_{\ell_2} \leq R_{\ell-1}$ and hence
\begin{displaymath}
    \frac{1}{R_\ell} 
    =
    \frac{2^n + \ell - n - 2^{\ell}}{2^n + \ell - n - 1}
    <
    \frac{2^n + \ell - n - 2^{\ell}}{2^n + \ell - n - 2}
    = 
    \frac{2}{R_{\ell - 1}} - \frac{1}{R_0}    
    \leq 
    \frac{1}{R_{\ell_1}} + \frac{1}{R_{\ell_2}} - \frac{1}{R_0}.   
\end{displaymath}
Let $c$ be the least common multiple of $R_0, R_1, \ldots, R_{n-1}$ and define $L_\ell = c R_\ell$. 
Then $c = L_0 < L_1 < \cdots < L_{n-1} = 2 c$ is an increasing sequence of integers.  
Note that the property \eqref{eq:thm:unique_optimal:ineq} holds also when $R$ is replaced by $L$ since the $c$ cancels out. 

\emph{Step 2. Construct an instance $\boldsymbol{k} = (k_0, k_1, \ldots, k_n)$.}
To make $\mathcal{A}$ uniquely optimal, we assign integers from the sequence $L_0, L_1, \ldots, L_{n-1}$ to the elements of $\boldsymbol{k}$ such that Lemma~\ref{thm:chin} guarantees that every matrix multiplication in $\mathcal{A}$ appears in every optimal parenthesisation.
In other words, $\mathcal A$ will be uniquely optimal on $\boldsymbol k$.

There must be some $i \in \{1, 2, \ldots, n - 1 \}$ such that $(i-1, i, i+1) \in \mathcal A$. 
This triplet corresponds to the multiplication $M_i M_{i+1}$. 
Set the dimension $k_i = L_{n-1}$.
Replace $M_i M_{i+1}$ by $\hat M_i := M_i M_{i+1}$ of size $k_{i-1} \times k_{i+1}$ and obtain a reduced chain of length $n-1$.
Recursively apply this reduction until only one matrix remains. 

In the first level of recursion, $L_{n-1}$ is assigned to some $k_{i_1}$. 
In the second level of recursion, $L_{n-2}$ is assigned to some $k_{i_2}$. 
Finally, in the last level of recursion, $L_1$ is assigned to some $k_{i_{n-1}}$.
At this point, each interior dimension $k_1, k_2, \ldots, k_{n-1}$ has been assigned an integer from the sequence $L_1, \ldots, L_{n-1}$. 
As the final step, set the exterior dimensions $k_0, k_n = L_0$.  
Since the elements of $L$ are assigned from largest to smallest, the inequality \eqref{eq:thm:unique_optimal:ineq} will hold for every matrix multiplication in $\mathcal{A}$.
Therefore, each multiplication in $\mathcal A$ must, by Lemma~\ref{thm:chin}, be present in every optimal parenthesisation. 
Since there is only one such parenthesisation, $\mathcal A$ must be uniquely optimal on $\boldsymbol k$.

\emph{Step 3. Go from one to infinitely many instances.}
By Lemma~\ref{lemma::infinity}, $\mathcal A$ is uniquely optimal on infinitely many instances. \qed
\end{proof}

\section{Few parenthesisations are essential}\label{sec:essential_algs}

Having established that all parenthesisations are useful, it is natural to ask whether they are \emph{equally} useful.
The answer is no. For a chain of length $n \ge 4$, 
there is a set of $n + 1$ parenthesisations that are far more useful than the others. 

\begin{definition}[Essential parenthesisation]\label{def:essential}
A parenthesisation $\mathcal{A}$ is \emph{essential} if for any threshold $r \in \mathbb R^+$, there exist infinitely many instances for which the removal of $\mathcal{A}$ induces a penalty that exceeds $r$. 
\end{definition}

In other words, there is no upper bound on the penalty when an essential parenthesisation is removed.
Conversely, removing a \emph{non-}essential parenthesisation causes a bounded penalty.

We now characterise the essential parenthesisations.
For any $h \in \{ 0, 1, \ldots, n \}$, let $\mathcal E_n^h \in \mathcal S_n$ denote the parenthesisation 
\begin{equation*}
(M_1 (M_2 \cdots (M_{h-1} M_h) \cdots ))(( \cdots (M_{h+1} M_{h+2})\cdots M_{n-1}) M_n).
\end{equation*}
In other words, (i) the prefix $M_1 \cdots M_h$ is computed right-to-left, (ii) the suffix $M_{h+1} \cdots M_n$ is computed left-to-right, (iii) once prefix and suffix have been computed, the two partial products are multiplied together. 
Notice that if $h \in \{0, n\}$, then either the prefix or the suffix becomes the entire matrix chain and two of the three stages are omitted.
Let us define the subset
\begin{equation*}
    \mathcal E_n = \{ \mathcal E_n^0, \mathcal E_n^1, \ldots, \mathcal E_n^n \} \subseteq \mathcal S_n
\end{equation*}
of parenthesisations.
The goal of this section is to prove that $\mathcal E_n$ contains all and only the essential parenthesisations.
In other words, $\mathcal E_n$ is \emph{the} set of essential parenthesisations.
Note that the size of $\mathcal E_n$ is $n+1$ for $n\ge 4$, while due to duplicates it is only $n-1$ for $n \in \{2, 3\}$. 
We begin by showing that all parenthesisations in $\mathcal E_n$ are essential.
\rebut{
The proof proceeds by demonstrating that removing any one of the parenthesisations in $\mathcal E_n$ leads to an unbounded penalty.
Therefore, all of them are essential. 
}
\begin{lemma}\label{lemma::essentials}
Consider a matrix chain of length~$n$. 
Then for any $h \in \{ 0, 1, \ldots, n \}$ and any threshold $r \in \mathbb{R}^{+}$, there exists an instance $\boldsymbol{k}$ such that $P(\{\mathcal{E}_n^h\}, \boldsymbol{k}) > r$.
Hence, by Lemma~\ref{lemma::infinity} all parenthesisations in $\mathcal E_n$ are essential. 
\end{lemma}
\begin{proof}
Consider an instance of the form $\boldsymbol k = (\alpha, \ldots, \alpha, 1, \alpha, \ldots, \alpha)$, where $\alpha > 1$ is an integer and $k_h = 1$.
The cost of $\mathcal{E}_n^h$ on $\boldsymbol{k}$ is
\begin{displaymath}
T(\mathcal{E}_n^h,\boldsymbol{k}) =
\sum_{1}^{n-1} 1 \cdot \alpha \cdot \alpha = (n-1) \alpha^{2}.
\end{displaymath}
Only $\mathcal E_n^h$ has $k_h$ appearing in each of the $n-1$ terms of the cost function. 
Any other parenthesisation has somewhere between $1$ and $n - 2$ terms in its cost function containing $k_h$.
Since $\alpha > 1$, the cost is minimised by maximising the number of terms containing $k_h$.
Therefore, let $\mathcal{B} \neq \mathcal{E}_n^h$ be a parenthesisation where $k_h$ appears in $n - 2$ terms of the cost function.
Its cost on $\boldsymbol{k}$ is
\begin{displaymath}
T(\mathcal{B},\boldsymbol{k}) =
\alpha\cdot\alpha\cdot\alpha + \sum_{1}^{n-2} 1\cdot\alpha\cdot\alpha = \alpha^{3} + (n-2) \alpha^{2}.
\end{displaymath}
Such a parenthesisation always exists. 
The penalty of removing $\mathcal{E}_n^h$ on $\boldsymbol k$ is
\begin{equation*}
\begin{split}
P(\{\mathcal{E}_n^h\}, \boldsymbol{k}) & =
\frac{\min_{\mathcal{A} \in \mathcal{S}_{n}\setminus\{\mathcal{E}_n^h\}}T(\mathcal{A}, \boldsymbol{k})}{T(\mathcal{E}_n^h, \boldsymbol{k})} - 1 =
\frac{T(\mathcal{B}, \boldsymbol{k})}{T(\mathcal{E}_n^h, \boldsymbol{k})} - 1 \\
& = \frac{\alpha^{3} + (n-2) \alpha^{2}}{(n-1) \alpha^{2}} - 1 = 
\frac{\alpha}{n-1} + \frac{n-2}{n-1} - 1 = \Theta(\alpha).
\end{split}
\end{equation*}
Since the penalty grows without bound as $\alpha \to \infty$, for any $r \in \mathbb R^+$ we can choose $\alpha$ large enough to make the penalty greater than $r$.
By Lemma~\ref{lemma::infinity}, there are infinitely many instances with the same penalty. \qed
\end{proof}

\rebut{Consider an instance $\boldsymbol{k}$ and choose any $m$ such that $k_m = \min_i k_i$.}
The next Lemma shows that the parenthesisation $\mathcal E_n^m$ has a cost that is strictly less than twice the optimal cost.
This result first appeared in Chandra's technical report~\cite[Theorem~3]{chandra1975computing}. 
We improve on Chandra's exposition of the proof. 
\rebutt{
The proof consists of two steps.
First, we show that a conveniently chosen sum is less than twice the optimal cost, by mapping terms in the former to larger terms in the latter.
Then, we show that the cost of $\mathcal{E}_{n}^{m}$ is less than the conveniently chosen sum.
}
The result holds not only for the cost function in \eqref{eq:cost-function-phi} but for any cost function $\phi_{\boldsymbol{k}}$ that is monotonically non-decreasing and has rotational symmetry in its arguments, meaning $\phi_{\boldsymbol{k}}(a, b, c) = \phi_{\boldsymbol{k}}(b, c, a)$. 
\begin{lemma}\label{lemma::chandra_modified}
    Let $T_{\rm opt}$ be the cost of an optimal parenthesisation on instance $\boldsymbol{k}$, and let $m$ be any index such that $k_m = \min_i k_i$.
    Then $T(\mathcal{E}_n^m, \boldsymbol{k}) < 2 T_{\rm opt}$.
\end{lemma}

\begin{proof}
    The case $n \leq 2$ is trivial, so assume $n \geq 3$.
    Let $\mathcal A$ be a parenthesisation such that $T_{\rm opt} = T(\mathcal{A}, \boldsymbol{k})$.
    We show that $T(\mathcal{E}_n^m, \boldsymbol{k}) < D \leq 2 T_{\rm opt}$, where $D$ is the conveniently chosen sum
    \begin{displaymath}
        D := \sum_{i = 0}^n \phi_{\boldsymbol k} (i - 1, i, m).
    \end{displaymath}
    Here, and in the rest of the proof, all arguments to $\phi_{\boldsymbol k}$ are modulo $n + 1$.
    In particular, the term in $D$ for $i = 0$ is $\phi_{\boldsymbol k}(n, 0, m)$.
    We will show that $D \leq 2 T_{\rm opt}$ (Step~1) and then show that $T(\mathcal{E}_n^m, \boldsymbol{k}) < D$ (Step~2).
    We first make an observation.

    \emph{Observation 1.}
    Consider any parenthesisation. 
    Each matrix $M_i$ appears in one and only one multiplication.
    If $M_i$ appears as the left operand $M_i X$ for some $X = \prod_{j=i+1}^z M_j$, then the term $\phi_{\boldsymbol k}(i-1, i, z)$ appears in the cost function.
    In this case, the dimension $k_i$ is eliminated and hence no other term can contain the argument $i$. 
    If instead $M_i$ appears as the right operand $X M_i$ for some $X = \prod_{j=z}^{i-1} M_j$, then the term $\phi_{\boldsymbol k}(z, i-1, i)$ appears in the cost.
    In this case, the dimension $k_{i-1}$ is eliminated and hence no other term can contain the argument $i - 1$. 
    In conclusion, there will be one and only one term in the cost function that contains both $i - 1$ and $i$ as arguments to $\phi_{\boldsymbol{k}}$ for all $i \in \{1, 2, \ldots, n\}$.
    Moreover, the final multiplication in any parenthesisation takes the form $(M_1 \cdots M_z) (M_{z+1} \cdots M_n)$ for some $1 \leq z \leq n - 1$.
    This multiplication adds the term $\phi_{\boldsymbol k}(0,z,n) = \phi_{\boldsymbol k}(n, 0, z)$ (equality due to rotational symmetry) to the cost.
    As a result, for each $i \in \{ 0, 1, \ldots, n \}$, $T_{\rm opt}$ has one and only one term of the form $\phi_{\boldsymbol k}(i-1,i,z)$ for some $z$.

    \emph{Step 1. Show that $D \leq 2 T_{\rm opt}$.}
    We will pair every term in $D$ with some term in $2 T_{\rm opt}$ such that each term in $D$ is less than or equal to its paired term in $2 T_{\rm opt}$.
    From this we can then conclude that $D \leq 2 T_{\rm opt}$.
    
    Let $\sigma$ be the function that for each $i \in \{ 0, 1, \ldots, n \}$ maps the term  $\phi_{\boldsymbol{k}}(i-1, i, m)$ in $D$ to a term of the form $\phi_{\boldsymbol k}(i-1,i,z)$ in $T_{\rm opt}$.
    This function is well-defined, since Observation~1 ensures that there is one and only one term of the specified form in $T_{\rm opt}$.
    
    Consider any term $t$ in $T_{\rm opt}$. 
    There are three cases: 
    \begin{itemize}
        \item $\sigma$ maps \emph{none} of the terms in $D$ to $t$.
        This happens if and only if there is no pair of consecutive indices in $t$. 
        That is, if $t = \phi_{\boldsymbol k}(a, b, c)$ then $a + 1 \not \equiv b \pmod{n+1}$, $b + 1 \not \equiv c \pmod{n + 1}$, and $c + 1 \not \equiv a \pmod{n + 1}$.
        
        \item $\sigma$ maps \emph{one} term in $D$ to $t$.
        This happens if and only if there is exactly one consecutive pair of indices in $t$.
        
        \item $\sigma$ maps \emph{two} terms in $D$ to $t$.
        This happens if and only if $t$ is of the form $\phi_{\boldsymbol{k}}(i-1, i, i+1)$ for some $i$ (modulo $n + 1$).
        
    \end{itemize}
    The function $\sigma$ cannot map \emph{more} than two terms in $D$ to $t$ when $n \geq 3$ since the system of congruences $a + 1 \equiv b \pmod{n+1}$, $b + 1\equiv c \pmod{n+1}$, and $c + 1 \equiv a \pmod{n+1}$ does not have a solution when $n \geq 3$.\footnote{If $n = 2$, then $a,b,c = 0,1,2$ satisfies the conditions and there are \emph{three} pairs of consecutive indices: $(0,1)$, $(1,2)$, and $(0,2)$.}
    Hence, $\sigma$ maps \emph{at most} two terms in $D$ to any given term in $T_{\rm opt}$.
    Since $\sigma$ is well-defined, every term in $D$ is mapped to some term in $T_{\rm opt}$, but not every term in $T_{\rm opt}$ is necessarily mapped to by any term in $D$.
    
    Since $2 T_{\rm opt} = T_{\rm opt} + T_{\rm opt}$ has two copies of each term and $\sigma$ never maps \emph{more} than two terms in $D$ to the same term in $T_{\rm opt}$, each term in $D$ can be paired with its own unique term in $2 T_{\rm opt}$ such that no term in the latter is paired up more than once.
    Each pair established in this way has the form $\phi_{\boldsymbol k}(i-1,i,m)$ in $D$ and $\phi_{\boldsymbol k}(i-1,i,z)$ in $T_{\rm opt}$.
    Since $k_m$ is the smallest element in $\boldsymbol k$ and the function $\phi_{\boldsymbol{k}}$ is monotonically non-decreasing in each argument, $\phi_{\boldsymbol k}(i-1,i,m) \leq \phi_{\boldsymbol k}(i-1,i,z)$.
    Therefore, $D \leq 2 T_{\rm opt}$.

    \emph{Step 2. Show that $T(\mathcal{E}_n^m, \boldsymbol{k}) < D$.}
    The cost of $\mathcal{E}_n^m$ on $\boldsymbol k$ can be expressed as
    \begin{displaymath}
        T(\mathcal{E}_n^m, \boldsymbol{k})= 
        \sum_{i \in \{ 0, 1, \ldots, n \} \setminus \{ m, m + 1 \} } \phi_{\boldsymbol k}(i - 1, i, m).
    \end{displaymath}
    Written this way, it is clear that every term in $T(\mathcal{E}_n^m, \boldsymbol{k})$ also appears in $D$.
    But since $D$ has strictly more terms, $T(\mathcal{E}_n^m, \boldsymbol{k}) < D$.

    By combining Steps~1 and~2, it follows that $T(\mathcal{E}_n^m, \boldsymbol{k}) < 2 T_{\rm opt}$. \qed
\end{proof}

The following example serves to illustrate the logic of the proof. 
Figure~\ref{fig:chandra-proof-mapping-illustration} illustrates the two layers of pairing of terms for $m = 3$ and the optimal parenthesisation
\begin{displaymath}
((M_1 M_2) ((M_3 (M_4 M_5)) M_6)) (M_7 M_8).
\end{displaymath}
The terms of $2 T_{\rm opt}$ are displayed on the top row; the terms of $D$ are displayed on the middle row; and the terms of $T(\mathcal{E}_n^m, \boldsymbol{k})$ are displayed on the bottom row. 
Every term of $D$ has been paired by $\sigma$ with a term of $2 T_{\rm opt}$ that is greater or equal to it.
Similarly, every term of $T(\mathcal{E}_n^m, \boldsymbol k)$ has been paired with its identical term in $D$.

\begin{figure}[htbp]
        \centering
        \begin{tikzpicture}[y=-3cm,x=1.46cm]
        \node [fill=gray!30] (O1) at (1,0) {$(0,1,2)$};
        \node (O2) at (2,0) {$(0,2,6)$};
        \node (O3) at (3,0) {$(2,3,5)$};
        \node (O4) at (4,0) {$(3,4,5)$};
        \node [fill=gray!30] (O5) at (6,0) {$(2,5,6)$};
        \node [fill=gray!30] (O6) at (0,0) {$(0,6,8)$};
        \node [fill=gray!30] (O7) at (7,0) {$(6,7,8)$};
        \node [fill=gray!30] (O1b) at (1.4,-0.2) {$(0,1,2)$};
        \node (O2b) at (2.4,-0.2) {$(0,2,6)$};
        \node (O3b) at (3.4,-0.2) {$(2,3,5)$};
        \node [fill=gray!30] (O4b) at (4.4,-0.2) {$(3,4,5)$};
        \node (O5b) at (6.4,-0.2) {$(2,5,6)$};
        \node (O6b) at (0.4,-0.2) {$(0,6,8)$};
        \node [fill=gray!30] (O7b) at (7.4,-0.2) {$(6,7,8)$};

        \node [align=center, fill=gray!30] (M1) at (0,1) {$(8,0,3)$\\$i=0$};
        \node [align=center, fill=gray!30] (M2) at (1,1) {$(0,1,3)$\\$i=1$};
        \node [align=center, fill=gray!30] (M3) at (2,1) {$(1,2,3)$\\$i=2$};
        \node [align=center              ] (M4) at (3,1) {$(2,3,3)$\\$i=3$};
        \node [align=center              ] (M5) at (4,1) {$(3,4,3)$\\$i=4$};
        \node [align=center, fill=gray!30] (M6) at (5,1) {$(4,5,3)$\\$i=5$};
        \node [align=center, fill=gray!30] (M7) at (6,1) {$(5,6,3)$\\$i=6$};
        \node [align=center, fill=gray!30] (M8) at (7,1) {$(6,7,3)$\\$i=7$};
        \node [align=center, fill=gray!30] (M9) at (8,1) {$(7,8,3)$\\$i=8$};

        \draw [->] (M1.north) -- (O6.south) node [midway, fill=white] {$\leq$};
        \draw [->] (M2.north) -- (O1.south)  node [midway, fill=white] {$\leq$};
        \draw [->] (M3.north) -- (O1b.south) node [midway, fill=white] {$\leq$};
        \draw [->] (M4.north) -- (O3.south)  node [midway, fill=white] {$\leq$};
        \draw [->] (M5.north) -- (O4.south)  node [midway, fill=white] {$\leq$};
        \draw [->] (M6.north) -- (O4b.south) node [midway, fill=white] {$\leq$};
        \draw [->] (M7.north) -- (O5.south)  node [midway, fill=white] {$\leq$};
        \draw [->] (M8.north) -- (O7.south)  node [midway, fill=white] {$\leq$};
        \draw [->] (M9.north) -- (O7b.south) node [midway, fill=white] {$\leq$};

        \node [fill=gray!30] (07m) at (0,2) {$(0,3,8)$};
        \node [fill=gray!30] (01m) at (1,2) {$(0,1,3)$};
        \node [fill=gray!30] (02m) at (2,2) {$(1,2,3)$};
        \node [fill=gray!30] (03m) at (5,2) {$(3,4,5)$};
        \node [fill=gray!30] (04m) at (6,2) {$(3,5,6)$};
        \node [fill=gray!30] (05m) at (7,2) {$(3,6,7)$};
        \node [fill=gray!30] (06m) at (8,2) {$(3,7,8)$};

        \draw [->] (01m.north) -- (M2.south) node [midway, fill=white] {$\leq$};
        \draw [->] (02m.north) -- (M3.south) node [midway, fill=white] {$\leq$};
        \draw [->] (03m.north) -- (M6.south) node [midway, fill=white] {$\leq$};
        \draw [->] (04m.north) -- (M7.south) node [midway, fill=white] {$\leq$};
        \draw [->] (05m.north) -- (M8.south) node [midway, fill=white] {$\leq$};
        \draw [->] (06m.north) -- (M9.south) node [midway, fill=white] {$\leq$};
        \draw [->] (07m.north) -- (M1.south) node [midway, fill=white] {$\leq$};
        
        \end{tikzpicture}
        \caption{Example of term-pairing between $2 T_{\rm opt}$ (top), $D$ (middle), and $T(\mathcal{E}_n^m, \boldsymbol{k})$ (bottom).}
        \label{fig:chandra-proof-mapping-illustration}
\end{figure}

Lemma~\ref{lemma::chandra_modified} implies that every parenthesisation $\mathcal A \not \in \mathcal E_n$ is \emph{non-}essential.
Suppose the contrary: $\mathcal A \in \mathcal S_n \setminus \mathcal E_n$ and is essential. 
Then by Definition~\ref{def:essential} there exist instances for which $\mathcal E_n^m$ is more than a factor of $2$ from optimal, thus contradicting Lemma~\ref{lemma::chandra_modified}.
The following theorem expands on this conclusion and establishes an upper bound on the penalty of removing every parenthesisation not in $\mathcal E_n$.

\begin{theorem}\label{theorem::non-essential}
The maximum penalty of removing every parenthesisation not in $\mathcal{E}_n$ is less than $100\%$, i.e., $P(\mathcal{S}_n \setminus \mathcal{E}_n) < 100\%$.
\end{theorem}
\begin{proof}
\rebutt{
Let $\boldsymbol{k}$ be any instance; choose any $m$ such that $k_m = \min_i k_i$.
From (\ref{eq:max_penalty_set}), the penalty of removing every parenthesisation not in $\mathcal{E}_n$ is
\begin{align*}
    P(\mathcal S_n \setminus \mathcal E_n) & = 
    \sup_{\boldsymbol{k} \in \mathbb{N}_{+}^{n+1}} \frac{\min_{\mathcal B \in \mathcal S_n \setminus (\mathcal S_n \setminus \mathcal E_n)} T(\mathcal B, \boldsymbol k)}{\min_{\mathcal A \in \mathcal S_n} T(\mathcal A, \boldsymbol k)} - 1 &&& \\
    & = 
    \sup_{\boldsymbol{k} \in \mathbb{N}_{+}^{n+1}} \frac{\min_{\mathcal B \in \mathcal E_n} T(\mathcal B, \boldsymbol k)}{\min_{\mathcal A \in \mathcal S_n} T(\mathcal A, \boldsymbol k)} - 1 &&& (\mathcal{E}_{n}^{m} \in \mathcal{E}_n) \\
    & \leq  \sup_{\boldsymbol{k} \in \mathbb{N}_{+}^{n+1}} \frac{T(\mathcal E_n^m, \boldsymbol k)}{\min_{\mathcal A \in \mathcal S_n} T(\mathcal A, \boldsymbol k)} - 1 <
    2 - 1 = 100\%, &&&
\end{align*}
}
where the final inequality follows from Lemma~\ref{lemma::chandra_modified}. \qed
\end{proof}

Combining Lemma~\ref{lemma::essentials} with Theorem~\ref{theorem::non-essential} proves our main result: The set $\mathcal E_n$ contains all and only essential parenthesisations.

\begin{theorem}
\label{thm:the-set-of-essential-algorithms}
All parenthesisations in $\mathcal{E}_n$ are essential and no other parenthesisation is essential. 
\end{theorem}
\begin{proof}
By Lemma~\ref{lemma::essentials}, all parenthesisations in $\mathcal{E}_n$ are essential.
By Theorem~\ref{theorem::non-essential}, no parenthesisation in $\mathcal{S}_n \setminus \mathcal E_n$ is essential.
\qed
\end{proof}

In summary, out of the $C_{n-1}$ parenthesisations only at most $n + 1$ play a more important role than the others. 
Removing any of the essential parenthesisations causes an unbounded penalty, whereas removing all the non-essential ones causes a bounded penalty of at most $100\%$.

\section{Approximation algorithms}\label{sec:approximation-algorithms}

In 1975, Chandra presented an $\Theta(n)$ approximation algorithm that returns a parenthesisation that is never more than twice as expensive as an optimal one~\cite[Algorithm~3]{chandra1975computing}. Notably, Chandra's algorithm returns an essential parenthesisation $\mathcal E_n^m$, where $k_m = \min_i k_i$.

The connection between Chandra's algorithm and the essential parenthesisations suggests an alternative algorithm that maintains the $\Theta(n)$ time complexity but returns a parenthesisation that is always at least as good as the one returned by Chandra.
Namely, select from the set of essential parenthesisation one that minimises the cost.




\begin{algorithm}[htbp]
    \DontPrintSemicolon 
    \caption{
        $h = \mathtt{MinEssential}(k_0, k_1, \ldots, k_n)$
    }\label{alg:min-essential}
    \KwIn{An instance $\boldsymbol{k} = (k_0, k_1, \ldots, k_n)$ of the MCP.}
    \KwOut{The index $h$ of an essential parenthesisation $\mathcal E_n^h$ with minimal cost.}
    $z_0 \gets k_0 k_n$\;
    \For{$i \gets 1, 2, \ldots, n$}{
        $z_i \gets k_{i-1} k_i$\;
    }
    $z \gets \sum_{i=0}^n z_i$\;
    \For{$i \gets 0, 1, \ldots, n - 1$}{
        $t_i \gets k_i (z - z_{i} - z_{i+1})$\;
    }
    $t_n \gets k_n (z - z_{n} - z_{0})$\;
    $h \gets 0$\;
    \For{$i \gets 1, 2, \ldots, n$}{
        \If{$t_i < t_h$}{
            $h \gets i$\;
        }
    }
    
    \Return{$h$}
\end{algorithm}

Naively minimising the cost takes time $\Theta(n^2)$: The cost of each parenthesisation has $\Theta(n)$ terms to sum and there are $\Theta(n)$ parenthesisations for which to calculate the cost. 
Algorithm~\ref{alg:min-essential} finds an essential parenthesisation with minimal cost in $\Theta(n)$ time by recognising that the cost function of $\mathcal E_n^h$ can be expressed as $k_h$ multiplied by a sum of all but two of the $n + 1$ pairwise products $k_0 k_1, k_1 k_2, \ldots, k_{n-1} k_n, k_n k_0$.
The algorithm computes all the pairwise products ($z_i$) and their sum ($z$) in linear time.
Using those intermediate results, the algorithm is now able to compute the cost of each essential parenthesisation ($t_i$) and find a minimiser ($h$) in linear time.

In 1978,~\cite{chin1978n} improved on Chandra's approximation algorithm by adding a preprocessing step that eliminates many particularly bad cases. 
Chin proved that his algorithm has a better approximation factor. 
We realised that the pseudo-code of the algorithm in~\cite{chin1978n} contains an error;
we provide here a corrected and revised version of Chin's algorithm as Algorithm~\ref{alg:chin}.
This algorithm consists of four phases.
The first phase initialises the data structures.
The second phase scans the instance for internal dimensions (i.e., excluding $k_0$ and $k_n$) that satisfy the sufficient condition in Lemma~\ref{thm:chin}.
If a dimension satisfies the condition, then the corresponding pair of matrices are associated together and the chain length is reduced by one. 
The third phase does the same but from the two opposing ends of the chain.
Finally, in the fourth phase Chandra's algorithm is applied to the remaining reduced chain (if anything remains at this point).

\rebut{
The output of Chin's algorithm is a vector $v = (v_1, v_2, \ldots, v_{n-1})$, which is a permutation of the first $n-1$ integers.
Specifically, the entry $v_i$ indicates that the multiplication $i$ in the chain (numbered left to right) is the $v_i$'th performed.
As an example, the vector $v = (4, 1, 2, 5, 3)$ corresponds to the parenthesisation $(M_1 ((M_2 M_3) M_4)) (M_5 M_6)$.
}

\begin{algorithm}[htbp]
    \DontPrintSemicolon
    \caption{
        $(v_1, v_2, \ldots, v_{n-1}) = \mathtt{Chin}(k_0, k_1, \ldots, k_n)$
    }\label{alg:chin}
    \KwIn{The dimensions $(k_0, k_1, \ldots, k_n)$ of a matrix chain of length $n$.}
    \KwOut{A permutation $(v_1, v_2, \ldots, v_{n-1})$ of the integers $\{ 1, 2, \ldots, n - 1 \}$ specifying the ordering of the multiplications: $v_i$ indicates that the multiplication $i$ in the chain (numbered left to right) is the $v_i$'th performed.}
    \tcp{Initialise}
    $m \gets 0$\;
    \For{$i \gets 1, 2, \ldots, n$}{
        \If{$k_i < k_m$}{
            $m \gets i$\;
        }
    }
    
    $(r_0, r_1, \ldots, r_n) \gets (1/k_0, 1/k_1, \ldots, 1/k_n)$\;
    $(v_1, v_2, \ldots, v_{n-1}) \gets (0, 0, \ldots, 0)$\;
    Let $Q$ be an empty deque\;
    \tcp{Scan forward}
    $\mathtt{PushBack}(Q, 0)$\;
    $a \gets 1$\;
    \For{$i \gets 1, 2, \ldots, n - 1$}{
        $\mathtt{PushBack}(Q, i)$\;
        \While{$\mathtt{Size}(Q) \geq 2$ {\rm \bfseries and} $r_{\mathtt{Back}(Q, 0)} + r_m < r_{\mathtt{Back}(Q, 1)} + r_{i + 1}$}{
            $v_{\mathtt{Back}(Q,0)} \gets a$\;
            $a \gets a + 1$\;
            $\mathtt{PopBack}(Q)$\;
        }
    }
    $\mathtt{PushBack}(Q, n)$\;
    \tcp{Nibble at both ends}
    $b \gets n - 1$\;
    \While{$\mathtt{Size}(Q) \geq 3$}{
        \uIf{$r_{\mathtt{Back}(Q, 0)} + r_m < r_{\mathtt{Back}(Q, 1)} + r_{\mathtt{Front}(Q, 0)}$}{
            $\mathtt{PopBack}(Q)$\;
            $v_{\mathtt{Back}(Q, 0)} \gets b$\;
            $b \gets b - 1$\;
        }
        \uElseIf{$r_{\mathtt{Front}(Q, 0)} + r_m < r_{\mathtt{Front}(Q, 1)} + r_{\mathtt{Back}(Q, 0)}$}{
            $\mathtt{PopFront}(Q)$\;
            $v_{\mathtt{Front}(Q, 0)} \gets b$\;
            $b \gets b - 1$\;
        }
        \Else{
            exit loop\;
        }
    }
    \tcp{Associate out from smallest dimension}
    \For{$i \gets m - 1, m - 2, \ldots, \mathtt{Front}(Q,0) + 1, m + 1, m + 2, \ldots, \mathtt{Back}(Q,0) - 1$}{
        \If{$v_i = 0$}{
            $v_i \gets a$\;
            $a \gets a + 1$\;
        }
    }
    \tcp{Final multiplication, if any}
    \If{$m \neq 0, n$ {\rm \bfseries and} $v_m = 0$}{
        $v_m \gets a$\;
    }
    \Return{$(v_1, v_2, \ldots, v_{n-1})$}\;
\end{algorithm}

Recall that the output of Chandra's algorithm belongs to the small set of essential parenthesisations $\mathcal E_n$. 
Chin's algorithm can produce better approximations in large parts due to its ability to return  parenthesisations that are \emph{non-}essential.
In fact, the following result shows that Chin's algorithm can return \emph{any} parenthesisation. 
\begin{lemma}\label{lemma::chin_generate_all}
    For each parenthesisation $\mathcal A \in \mathcal S_n$, there exists an instance for which Algorithm~\ref{alg:chin} returns $\mathcal A$.
\end{lemma}
\begin{proof}
    In the proof of Theorem~\ref{theorem::unique_optimal}, we construct for each parenthesisation $\mathcal A \in \mathcal S_n$ an instance $\boldsymbol{k}$ such that $\mathcal A$ is optimal on that instance. 
    This construction is done in such a way that Chin's algorithm will reduce the entire chain during the preprocessing stage and then return precisely $\mathcal A$.
    The result follows since we can do this for any $\mathcal A$. 
    \qed
\end{proof}

Since Chin's algorithm applies Chandra's algorithm in its last stage, this algorithm can also be improved by minimising over the essential parenthesisations of the reduced chain instead of starting from a smallest dimension. 
See Algorithm~\ref{alg:chin-improved} for details; the only tricky part is translating indices to and from the reduced chain.
The first three stages are taken verbatim from Chin's algorithm; only the last phase is replaced.
Algorithm~\ref{alg:min-essential} is invoked to find an essential parenthesisation of minimum cost for the reduced chain. 

Since Algorithm~\ref{alg:chin-improved}'s first three phases are identical to Chin's, the difference (if any) between the parenthesisations selected by these two algorithms stems from the last phase.
Clearly, choosing the best essential parenthesisation is never worse than choosing the one that Chandra's algorithm returns. 
Therefore, Algorithm~\ref{alg:chin-improved} always yields a parenthesisation with a cost that is less than or equal to that of Chin's algorithm.

In fact, the approximation factor of Algorithm~\ref{alg:chin-improved} can be slightly improved. 
The approximation factor for Chin's algorithm proved by~\cite{hu1981n} is actually a series of factors: One for each matrix chain length $n$. 
The worst case is obtained for $n = 4$ and the factors get smaller as $n$ increases.
But for $n \leq 4$, \emph{all} parenthesisations are essential, meaning that our proposed improvement in Algorithm~\ref{alg:chin-improved} produces an optimal solution for $n \leq 4$.
It follows that Hu and Shing's bound for $n = 5$ can be used as an upper bound for Algorithm~\ref{alg:chin-improved}.
\rebutt{
Let $A : \mathbb{N}^{n + 1} \rightarrow \mathcal{S}_n$ be an approximation algorithm and $T_{\rm{opt}}$ be the cost of an optimal parenthesisation.
Then, the increase in cost relative to optimal of the parenthesisation returned by some approximation algorithm on some instance $\boldsymbol{k}$ can be expressed as 
\begin{equation}\label{eq:increase_cost}
    I(A, \boldsymbol{k}) = T(A(\boldsymbol{k}), \boldsymbol{k}) / T_{\rm{opt}}(\boldsymbol{k}) - 1,
\end{equation}
Thus, $\max_{\boldsymbol{k}} I(A, \boldsymbol{k}) + 1$ is the approximation factor of an approximation algorithm $A$.
}
\begin{lemma}
    \rebutt{
    Let Algorithm~\ref{alg:chin-improved} be denoted as $A_{3}$.
    The approximation factor for Algorithm~\ref{alg:chin-improved} is $\max_{\boldsymbol{k}} I(A_{3}, \boldsymbol{k}) + 1 < 1.1429$.
    }
\end{lemma}
\begin{proof}
    For $n \leq 3$, Chin's algorithm gives the optimal solution.
    Hence, we can assume $n \geq 4$.
    Theorem 8 in~\cite{hu1981n} shows that Chin's Algorithm's error ratio $R$ is a function of $n$, given by
    \begin{equation*}
        \label{eq:hs-factor}
        R(n) = \frac{t - 1}{t^2 + t + (n-3)},
    \end{equation*}
    where $t = 1 + \sqrt{n - 1}$.
    This error ratio is inversely proportional to $n$ and maximum for $n=4$ and $t = 1 + \sqrt{3}$.
    All parenthesisations are essential for $n=4$, so Algorithm~\ref{alg:chin-improved} returns an optimal solution.
    \rebutt{
    Therefore, Algorithm~\ref{alg:chin-improved}'s approximation factor can be more tightly bounded by $\max_{\boldsymbol{k}} I(A_{3}, \boldsymbol{k}) + 1 \leq R(5) + 1 = 1/7 + 1 < 1.1429$.
    }
    \qed
\end{proof}


\rebut{
Chin's algorithm approaches the theoretical bound on
reduced chains of the form $(x, x, px, px, x)$, with $p = 1 + \sqrt{3}$ (see~\cite{chin1978n}). 
By contrast, Algorithm~\ref{alg:chin-improved} is optimal on such instances.
This observation, combined with the results shown in Table~\ref{tab:algorithms-max-penalty}, leads us to conjecture that the approximation factor of Algorithm~\ref{alg:chin-improved} is not tight.
}

\begin{algorithm}[htbp]
    \DontPrintSemicolon
    \caption{
        $(v_1, v_2, \ldots, v_{n-1}) = \mathtt{ReduceAndMin}(k_0, k_1, \ldots, k_n)$
    }\label{alg:chin-improved}
    \KwIn{See Algorithm~\ref{alg:chin}.}
    \KwOut{See Algorithm~\ref{alg:chin}.}
    \tcp{Initialise}
    ...\;
    \tcp{Scan forward}
    ...\;
    \tcp{Nibble at both ends}
    ...\;
    \tcp{Minimise over the essential parenthesisations of the reduced chain}
    \If{$\mathtt{Size}(Q) \geq 3$}{
        \For{$i \gets 0, 1, \ldots, \mathtt{Size}(Q) - 1$}{
            $q_i \gets \mathtt{Front}(Q,i)$\;
        }
        $h_q = \mathtt{MinEssential}(k_{q_00}, k_{q_1}, \ldots, k_{q_{\mathtt{Size}(Q)-1}})$\;
        $h \gets \mathtt{Front}(Q, h_q)$\;
        \tcp{Apply the selected essential parenthesisation}
        \For{$i \gets h - 1, h - 2, \ldots, \mathtt{Front}(Q,0) + 1, h + 1, h + 2, \ldots, \mathtt{Back}(Q,0) - 1$}{
            \If{$v_i = 0$}{
                $v_i \gets a$\;
                $a \gets a + 1$\;
            }
        }
        \tcp{Final multiplication, if any}
        \If{$h \neq 0, n$ {\rm \bfseries and} $v_h = 0$}{
        $v_h \gets a$\;
        }
    }
    \Return{$(v_1, v_2, \ldots, v_{n-1})$}\;
\end{algorithm}

In summary, we discussed two pairs of approximation algorithms.
The first pair consists of Chandra's algorithm and our improved Algorithm~\ref{alg:min-essential}. 
These algorithms are simple to implement and select one of the $n + 1$ essential parenthesisations.
These parenthesisations are particularly simple to evaluate, since they break down into a left-to-right and a right-to-left partial product (or just one of these two).
Their main disadvantage is the larger theoretical approximation factor (a maximum penalty of $100\%$).

The second pair consists of Chin's algorithm and our improved version, Algorithm~\ref{alg:chin-improved}.
These algorithms are more convoluted and select from the entire set of parenthesisations, $\mathcal{S}_n$.
The more complicated structure of the non-essential parenthesisations also makes them more difficult to evaluate.
Their main advantage is the smaller theoretical approximation factor (a maximum penalty of $15.5\%$ and $14.3\%$, respectively). 

\section{Experiments}
The difference in approximation factor across the various algorithms is significant but somewhat misleading, since the worst cases captured by the approximation factors do not say anything about the expected accuracy in practice on random instances. 
We empirically investigated the average accuracy for short matrix chains with the following experiment\footnote{Experiments can be reproduced with the code in \href{https://github.com/HPAC/essential}{https://github.com/HPAC/essential}.}.

\rebutt{
For each $n \in \{4,5,\ldots, 10\}$, we randomly sampled one million instances 
of the MCP of length~$n$.
The dimensions were drawn uniformly at random from the interval $1 \leq k_i \leq 1000$. 
For each instance, we calculated the increase in cost relative to the optimal cost with (\ref{eq:increase_cost}) for each of the four approximation algorithms discussed in Section~\ref{sec:approximation-algorithms}.
We present three different metrics built on the increased cost: the average increase, the maximum increase, and the percentage of the instances for which there was a zero increase across the sampled instances.
}

\newcolumntype{d}[1]{D{.}{.}{#1}}

\begin{table}[htbp]
    \centering
    \caption{Average cost increase (percentage) over optimum for different $O(n)$ approximation algorithms. \rebutt{In bold, entries with the smallest increase for each $n$.}}
    \label{tab:algorithms-avg-penalty}
    \begin{tabular}{r | r@{.}l r@{.}l r@{.}l r@{.}l r@{.}l r@{.}l r@{.}l}
      \toprule
        $n$ & \multicolumn{2}{c}{$4$} & \multicolumn{2}{c}{$5$} & \multicolumn{2}{c}{$6$} & \multicolumn{2}{c}{$7$} & \multicolumn{2}{c}{$8$} & \multicolumn{2}{c}{$9$} & \multicolumn{2}{c}{$10$} \\
      \midrule
        Chandra & 0&59 & 0&55 & 0&48 & 0&41 & 0&36 & 0&31 & 0&27 \\
        Algorithm~\ref{alg:min-essential} & \textbf{0}&\textbf{0} & 0&30 & 0&35 & 0&34 & 0&31 & 0&28 & 0&24 \\
      \midrule
        Chin & 0&018 & 0&032 & 0&037 & 0&038 & 0&036 & 0&034 & 0&031 \\
        Algorithm~\ref{alg:chin-improved} & \textbf{0}&\textbf{0} & \textbf{0}&\textbf{00015} & \textbf{0}&\textbf{0024} & \textbf{0}&\textbf{0062} & \textbf{0}&\textbf{0098} & \textbf{0}&\textbf{013} & \textbf{0}&\textbf{015} \\
    \bottomrule
    \end{tabular}
\end{table}

Table~\ref{tab:algorithms-avg-penalty} shows the average increase in cost (as a percentage over optimal) of the parenthesisations returned by the approximation algorithms.
\rebut{
For instance, the average cost increase of Chandra's algorithm for $n = 5$ was $0.55\%$.
Similarly, Chin's algorithm had an average cost increase for $n = 6$ of $0.037\%$.
}
In general, all four approximation algorithms performed, on average, very close to optimal.

\rebutt{
For $n=4$, every parenthesisation is essential and both Algorithm~\ref{alg:min-essential} and Algorithm~\ref{alg:chin-improved} always find an optimal parenthesisation.
Chin's algorithm outperforms Chandra's and Algorithm~\ref{alg:min-essential}.
At the same time, as expected, these algorithms outperform their original counterparts regardless of the length of the chain.
However, the gap in increased cost over the optimum between the improved Algorithms~\ref{alg:min-essential} and~\ref{alg:chin-improved} and their corresponding counterparts appears to shrink as $n$ increases.
This suggest that the improvement on the original algorithms is most effective for short chains.
}

\begin{table}[htbp]
    \centering
    \caption{Maximum cost increase (percentage) over optimum for different approximation algorithms. \rebutt{In bold, entries with the lowest maximum increase for each $n$.}}
    \label{tab:algorithms-max-penalty}
    \begin{tabular}{r | r@{.}l r@{.}l r@{.}l r@{.}l r@{.}l r@{.}l r@{.}l}
      \toprule
        $n$ & \multicolumn{2}{c}{$4$} & \multicolumn{2}{c}{$5$} & \multicolumn{2}{c}{$6$} & \multicolumn{2}{c}{$7$} & \multicolumn{2}{c}{$8$} & \multicolumn{2}{c}{$9$} & \multicolumn{2}{c}{$10$} \\
      \midrule
        Chandra & 73&5 & 67&1 & 59&3 & 57&9 & 46&2 & 37&9 & 32&4 \\
        Algorithm~\ref{alg:min-essential} & \textbf{0}&\textbf{0} & 31&9 & 31&7 & 33&5 & 38&3 & 27&8 & 25&6 \\
      \midrule
        Chin & 14&8 & 12&7 & 11&6 & 9&9 & 11&0 & 9&8 & 8&5 \\
        Algorithm~\ref{alg:chin-improved} & \textbf{0}&\textbf{0} & \textbf{3}&\textbf{6} & \textbf{6}&\textbf{1} & \textbf{6}&\textbf{0} & \textbf{5}&\textbf{3} & \textbf{5}&\textbf{3} & \textbf{5}&\textbf{9} \\
    \bottomrule
    \end{tabular}
\end{table}

\rebutt{
The maximum cost increases were also recorded and shown in Table~\ref{tab:algorithms-max-penalty}.
For $n=4$, the maximum cost increase for Chandra's and Chin's algorithms approaches the theoretical bound.
In contrast, Algorithm~\ref{alg:min-essential} and Algorithm~\ref{alg:chin-improved} do not approach the theoretical bound, for any chain length.
More specifically, the maximum cost increase over optimum for Algorithm~\ref{alg:min-essential} and Algorithm~\ref{alg:chin-improved} is $38.3\%$ (for $n=8$) and $6.1\%$ (for $n=6$), respectively.
These results support our conjecture that the bound for Algorithm~\ref{alg:chin-improved} ($14.3\%$) is far from tight.
}

\begin{table}[htbp]
    \centering
    \caption{Percentage of the samples for which an optimal solution was found by some approximation algorithms. \rebutt{In bold, entries with the highest percentage for each $n$.}}
    \label{tab:algorithms-optimal}
    \begin{tabular}{r | r@{.}l r@{.}l r@{.}l r@{.}l r@{.}l r@{.}l r@{.}l}
      \toprule
        $n$ & \multicolumn{2}{c}{$4$} & \multicolumn{2}{c}{$5$} & \multicolumn{2}{c}{$6$} & \multicolumn{2}{c}{$7$} & \multicolumn{2}{c}{$8$} & \multicolumn{2}{c}{$9$} & \multicolumn{2}{c}{$10$} \\
      \midrule
        Chandra & 92&14 & 90&28 & 89&13 & 88&50 & 88&00 & 87&75 & 87&54 \\
        Algorithm~\ref{alg:min-essential} & \textbf{100}&\textbf{00} & 92&16 & 89&88 & 88&86 & 88&21 & 87&88 & 87&63 \\
      \midrule
        Chin & 99&44 & 98&81 & 98&27 & 97&85 & 97&57 & 97&34 & 97&21 \\
        Algorithm~\ref{alg:chin-improved} & \textbf{100}&\textbf{00} & \textbf{99}&\textbf{97} & \textbf{99}&\textbf{77} & \textbf{99}&\textbf{38} & \textbf{98}&\textbf{94} & \textbf{98}&\textbf{46} & \textbf{98}&\textbf{08} \\
    \bottomrule
    \end{tabular}
\end{table}

The approximation factors can be misleading for yet another reason.
For most instances, all of the approximation algorithms return an optimal solution.
Table~\ref{tab:algorithms-optimal} presents the fraction (as a percentage) of the sampled instances for which an optimal parenthesisation was found by the approximation algorithms. 
All algorithms found an optimal parenthesisation on most instances.
The two more convoluted algorithms failed to find an optimal solution in only a small percentage of the instances. 
The two improved algorithms find optimal solutions marginally more frequently than their counterparts. 
However, it is worth noting that some essential parenthesisation is optimal on most instances, given that Chandra's algorithm and Algorithm~\ref{alg:min-essential} select exclusively from the essential set.

In summary, all four approximation algorithms are, on average, optimal for most practical purposes.
\rebut{However, Chandra's algorithm and Algorithm~\ref{alg:min-essential} are simpler to implement than the rest.
For this reason, they might be preferred in all cases unless guaranteeing a better worst-case performance is crucial.}



\section{Conclusions}\label{sec::conclusions}

We studied the parenthesisations of the matrix chain multiplication problem (MCP). 
We found that all $C_{n-1}$ parenthesisations are useful, i.e., uniquely optimal somewhere in the problem domain. 
More significantly, we found that only $n + 1$ parenthesisations are essential, meaning that removing any one of them leads to an unbounded penalty on an infinite number of instances. 
We showed that the concept of essential parenthesisations is closely connected to the approximation algorithm presented in~\cite{chandra1975computing}.
We proposed an equally simple but improved approximation algorithm that chooses an essential parenthesisation with minimal cost.
We also proposed a similarly improved version of the algorithm given in~\cite{chin1978n}, which is the best known approximation algorithm to date.
\rebut{Our algorithm has an approximation factor of $0.143$, down from Chin's $0.155$, although we conjecture that this bound is not tight.}
We showed empirically that all four approximation algorithms are very close to optimal in almost all cases.

The importance of essential parenthesisations goes beyond the improvements to Chandra's and Chin's algorithms. 
In the context of automatic generation of algorithms and code for linear algebra expressions (see e.g., Linnea from~\cite{barthels2018generalized,barthels2021linnea}, BTO from~\cite{4536183}, and SLinGen from~\cite{Spampinato:18}), essential parenthesisations make it possible for a compiler (or library) to pre-generate a very limited set of candidates, whose cost can then be quickly evaluated at runtime.

%
%


%
%

\bibliographystyle{plainnat}      
\bibliography{references.bib}   

\begin{thebibliography}{13}
\providecommand{\natexlab}[1]{#1}
\providecommand{\url}[1]{\texttt{#1}}
\expandafter\ifx\csname urlstyle\endcsname\relax
  \providecommand{\doi}[1]{doi: #1}\else
  \providecommand{\doi}{doi: \begingroup \urlstyle{rm}\Url}\fi

\bibitem[Barthels et~al.(2018)Barthels, Copik, and Bientinesi]{barthels2018generalized}
Henrik Barthels, Marcin Copik, and Paolo Bientinesi.
\newblock The generalized matrix chain algorithm.
\newblock In \emph{Proceedings of the 2018 International Symposium on Code Generation and Optimization}, pages 138--148. ACM, 2018.

\bibitem[Barthels et~al.(2021)Barthels, Psarras, and Bientinesi]{barthels2021linnea}
Henrik Barthels, Christos Psarras, and Paolo Bientinesi.
\newblock Linnea: Automatic generation of efficient linear algebra programs.
\newblock \emph{ACM Transactions on Mathematical Software (TOMS)}, 47\penalty0 (3):\penalty0 1--26, 2021.

\bibitem[Chandra(1975)]{chandra1975computing}
A.K. Chandra.
\newblock Computing matrix chain products in near-optimal time.
\newblock Technical Report RC-5625 (\#24393), IBM Thomas J. Watson Research Center, P. O. Box 218, Yorktown Heights, New York, USA, 1975.

\bibitem[Chin(1978)]{chin1978n}
Francis~Y. Chin.
\newblock An {O}(n) algorithm for determining a near-optimal computation order of matrix chain products.
\newblock \emph{Communications of the ACM}, 21\penalty0 (7):\penalty0 544--549, 1978.

\bibitem[Godbole(1973)]{godbole1973efficient}
Sadashiva~S. Godbole.
\newblock On efficient computation of matrix chain products.
\newblock \emph{IEEE Transactions on Computers}, 100\penalty0 (9):\penalty0 864--866, 1973.

\bibitem[Hu and Shing(1981)]{hu1981n}
T.~C. Hu and M.~T. Shing.
\newblock An {O}(n) algorithm to find a near-optimum partition of a convex polygon.
\newblock \emph{Journal of Algorithms}, 2\penalty0 (2):\penalty0 122--138, 1981.

\bibitem[Hu and Shing(1982)]{hu1982computation}
T.~C. Hu and M.~T. Shing.
\newblock Computation of matrix chain products. {P}art {I}.
\newblock \emph{SIAM Journal on Computing}, 11\penalty0 (2):\penalty0 362--373, 1982.

\bibitem[Hu and Shing(1984)]{hu1984computation}
T.~C. Hu and M.~T. Shing.
\newblock Computation of matrix chain products. {P}art {II}.
\newblock \emph{SIAM Journal on Computing}, 13\penalty0 (2):\penalty0 228--251, 1984.

\bibitem[Ramanan(1994)]{Ramanan1994}
Prakash Ramanan.
\newblock A new lower bound technique and its application: Tight lower bound for a polygon triangulation problem.
\newblock \emph{SIAM Journal on Computing}, 23\penalty0 (4):\penalty0 834--851, 1994.
\newblock \doi{10.1137/S0097539790190077}.

\bibitem[Schwartz and Weiss(2019)]{schwartz2019revisiting}
Oded Schwartz and Elad Weiss.
\newblock Revisiting “computation of matrix chain products''.
\newblock \emph{SIAM Journal on Computing}, 48\penalty0 (5):\penalty0 1481--1486, 2019.

\bibitem[Siek et~al.(2008)Siek, Karlin, and Jessup]{4536183}
Jeremy~G. Siek, Ian Karlin, and E.~R. Jessup.
\newblock Build to order linear algebra kernels.
\newblock In \emph{2008 IEEE International Symposium on Parallel and Distributed Processing}, pages 1--8, 2008.
\newblock \doi{10.1109/IPDPS.2008.4536183}.

\bibitem[Spampinato et~al.(2018)Spampinato, Fabregat-Traver, Bientinesi, and P{\"u}schel]{Spampinato:18}
Daniele~G. Spampinato, Diego Fabregat-Traver, Paolo Bientinesi, and Markus P{\"u}schel.
\newblock Program generation for small-scale linear algebra applications.
\newblock In \emph{International Symposium on Code Generation and Optimization (CGO)}, pages 327--339, 2018.

\bibitem[Yao(1982)]{yao1982speed}
F.~Frances Yao.
\newblock Speed-up in dynamic programming.
\newblock \emph{SIAM Journal on Algebraic Discrete Methods}, 3\penalty0 (4):\penalty0 532--540, 1982.

\end{thebibliography}


\section*{Statements and Declarations}

\begin{itemize}
    \item \textbf{Funding:} Support by the Swedish Government’s strategic collaborative research programme "eSSENCE" is gratefully acknowledged.

    \item \textbf{Competing Interests:} All authors declare they have no financial interests.

    \item \textbf{Author Contributions:} All authors contributed to the conception and design of this study. Implementation and experimentation were performed by Francisco López, who also drafted a first version of the theoretical proofs and of the manuscript, which were then commented on and improved by all authors. Lars Karlsson developed sound versions of the proofs. Lars Karlsson and Paolo Bientinesi supervised the whole development and finalised the manuscript.

    \item \textbf{Data Availability:} Not applicable.
\end{itemize}

\end{document}